\documentclass[10pt,twocolumn,letterpaper]{article}

\usepackage[pagenumbers]{wacv} 

\usepackage{graphicx}
\usepackage{amsmath}
\usepackage{amssymb}
\usepackage{booktabs}

\usepackage[pagebackref,breaklinks,colorlinks]{hyperref}

\usepackage[capitalize]{cleveref}
\crefname{section}{Sec.}{Secs.}
\Crefname{section}{Section}{Sections}
\Crefname{table}{Table}{Tables}
\crefname{table}{Tab.}{Tabs.}

\def\wacvPaperID{1760} 
\def\confName{WACV}
\def\confYear{2025}

\begin{document}

\title{Towards Population Scale Testis Volume Segmentation in DIXON MRI}

\author{
Jan Ernsting\\
University of Münster\\
Germany\\
{\tt\small j.ernsting@uni-muenster.de}
\and
Phillip Nikolas Beeken\and 
Lynn Ogoniak\and 
Jacqueline Kockwelp\and
Tim Hahn\and 
Alexander Siegfried Busch\and 
Benjamin Risse\\ 
University of Münster\\
Germany\\
{\tt\small b.risse@uni-muenster.de}
}
\maketitle

\begin{abstract}
  Testis size is known to be one of the main predictors of male fertility, usually assessed in clinical workup via palpation or imaging. 
  Despite its potential, population-level evaluation of testicular volume using imaging remains underexplored. 
  Previous studies, limited by small and biased datasets, have demonstrated the feasibility of machine learning for testis volume segmentation. 
  This paper presents an evaluation of segmentation methods for testicular volume using Magnet Resonance Imaging data from the UKBiobank. 
  The best model achieves a median dice score of $0.87$, compared to median dice score of $0.83$ for human interrater reliability on the same dataset, enabling large-scale annotation on a population scale for the first time. 
  Our overall aim is to provide a trained model, comparative baseline methods, and annotated training data to enhance accessibility and reproducibility in testis MRI segmentation research.
\end{abstract}

\section{Introduction}
Reproductive health is a fundamental aspect of overall health, and is essential for the survival and prosperity of individuals and populations. 
Approximately one in ten men of reproductive age experiences fertility issues~\cite{KRAUSZ2011271}. 
Testicular volume is known to be the main predictor of total sperm count in men and thereby male fertility~\cite{Andrology4thed}. 
It is therefore routinely assessed in andrologic clinical workup using palpation or imaging.

Deep learning is a current technique of computer science allowing to apply machine learning techniques directly on high dimensional data, including for example imaging and natural language. 
Extracting insights from these forms of data, e.g. imaging data, has proven to be especially useful in the medical domain~\cite{mlinmedicine}.

Despite the availability of novel analysis techniques and applicability to large scale data, the systematic evaluation of testicular volume at population level using imaging data is still lacking. 
To our knowledge, only one previous study has used deep learning for segmentation of testicular volume in a retrospective study based on a cohort of male patients undergoing Magnet Resonance Imaging (MRI) for evaluation of a preexisting andrologic condition~\cite{sun_magnetic_2023}. 
While the study provided evidence for the feasibility of machine learning for testis volume segmentation, the sample was strongly biased towards patients undergoing advanced clinical evaluation for andrologic conditions, such as infertility or testicular pain, and was therefore non-representative of the general population.
Furthermore, the study focused on a relatively small and custom dataset comprising $200$ MRI T2 weighted images, hindering its applicability for large datasets such as the UKBiobank, one of the currently largest publicly available population samples conducted in the united kingdom, due to differences in MRI modalities.

\subsection{Contribution}
The aim of this paper is exploration of segmentation in testis MRI for subsequent analysis given a population sample. 
In order to investigate in this direction, we propose three core contributions. 
First, we provide a novel annotated dataset for training of segmentation models on the UKBiobank dataset. 
Second, we investigate the quality of annotations from human expert annotators by repeated annotation of the same annotator and annotation by a clinical radiologist. 
Finally, we publish the final model and the according weights to provide a baseline and to enable applications of our findings.

\subsection{Related Work}

\subsubsection{Segmentation} Segmentation is a vital area of research in the broader realm of computer vision and deep learning applications. 
Current models are most commonly convolution based architectures, for example DeepLabV3~\cite{deeplabv3} or more recent DeepLabV3Plus~\cite{deeplabv3plus}. 
Despite their success one of the most employed architectures in medical segmentation remains the popular U-Net~\cite{unet} architecture.
All above mentioned convolutional models are optimized for two-dimensional image data. 
Additionally, we included a 3D version of the U-Net architecture to compare performance of 3D models.


Recent advances in the area of transformer models~\cite{vaswani2023attentionneed, dosovitskiy2021imageworth16x16words} enabled application of attention in segmentation. 
Vision Transformer (ViT)~\cite{dosovitskiy2021imageworth16x16words} inspired backbones are becoming increasingly popular. 
To provide an extensive comparison of available methods, we include architectures with both convolutional and transformer backbones in our analysis. 
Due to the higher demand on available manually annotated training data, we focus on convolutional backbones.

\subsubsection{Segmentation of MRI data} MRI is a frequently and routinely used medical imaging technique in clinical context. 
One recent application of computer aided segmentation on MRI data of the UKBiobank is~\cite{ukbbkidney}. 
Beside those studies focused on population-based databases there exists a growing body of literature in medicine in benchmarks and competitions, for example the BRATS~\cite{Brats1, brats2}, LITS~\cite{Bilic2023}, PAC~\cite{Fisch2021} and others~\cite{Antonelli2022}.

\subsubsection{Evaluation of testicular volume} In the clinical routine several methods are used for the assessment of testicular volume as proxy for testicular function~\cite{lotti_european_2020}. 
Conventional methods include Prader orchidometer, a series of ellipsoid beads of increasing size, and ultrasonography, which represents the current clinical gold standard method\cite{lotti_european_2020}. 
However, both techniques have notable limitations. 
The Prader orchidometer tends to overestimate testicular volume, particularly in smaller testes, and depends on the clinician's visual and tactile assessment introducing substantial inter-observer variability~\cite{carlsen_inter-observer_2000}. 
Ultrasonography estimates, widely regarded as more accurate compared to Prader orchidometer, vary depending on the mathematical formula used and the lack of international standardization impedes the comparison between studies. 
The difference between these conventional methods results in a mean difference of approx. 5 milliliters~\cite{sakamoto_testicular_2007} leading to a substantial diagnostic uncertainty. 
MRI is rarely used in the initial andrologic evaluation of a patient, but plays an important role as secondary diagnostic, non-invasive imaging tool for testicular pathologies~\cite{parenti_imaging_2018}. 
Furthermore this technique allows the precise assessment of testicular volumes overcoming the limitations of conventional methods.
A recent study shows feasibility of application of machine learning for testis providing means to address the mentioned shortcomings using segmentation algorithms, namely the ResUnet approach~\cite{testismri}.

\section{Materials and Methods}

\subsection{Dataset}
\begin{figure}
\begin{subfigure}{\linewidth}
  \centering
  \includegraphics[width=.5\linewidth]{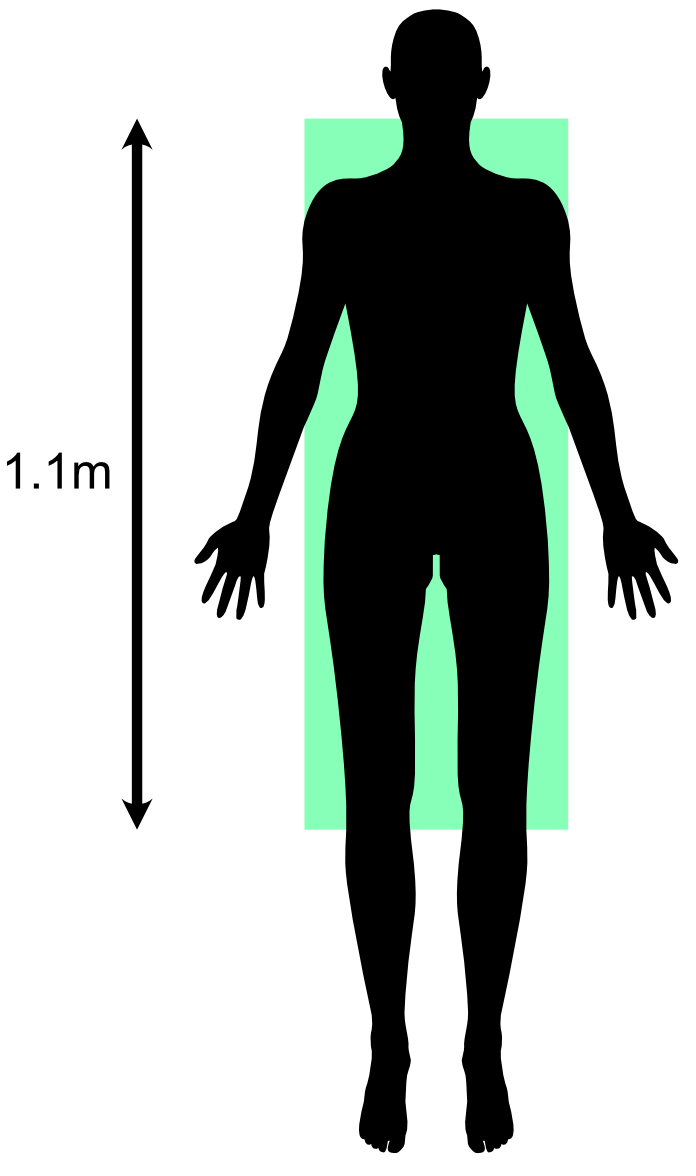}
  \caption{Measured area}
  \label{fig:sfig1}
\end{subfigure}
~
\begin{subfigure}{\linewidth}
  \centering
  \includegraphics[width=.5\linewidth]{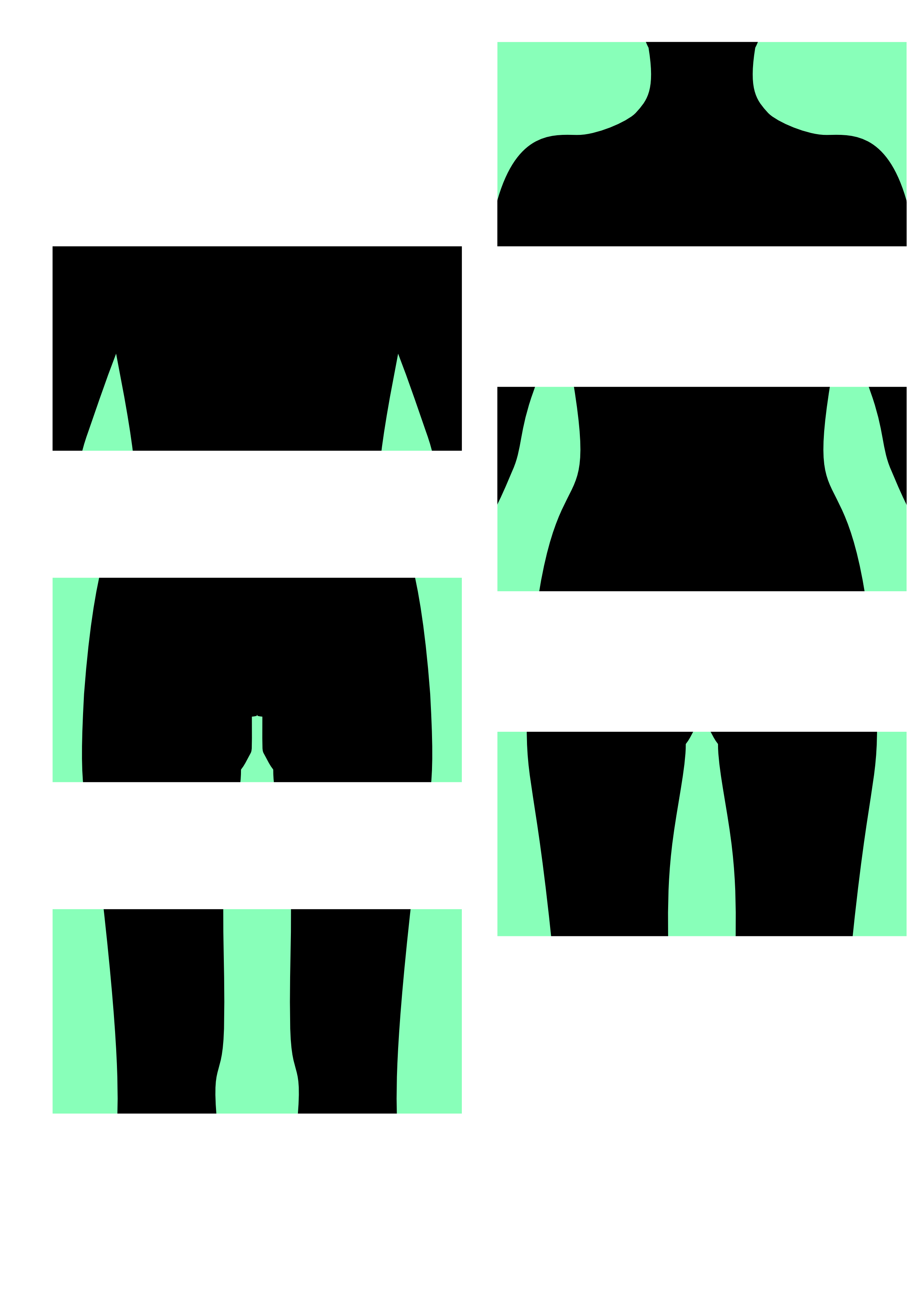}
  \caption{Imaging windows}
  \label{fig:sfig2}
\end{subfigure}
\caption{Sketch of UKBiobank DIXON measurements. From the 1.1 meters hight of measurement in the UKBiobank (a), a total of six overlapping imaging windows are extracted (b).}
\label{fig:ukbbmeasurements}
\end{figure}

We used the UKBiobank dataset, a publicly available population-based study conducted in the United Kingdom.
The study represents a large-scale biomedical research resource that contains in-depth genetic and health information from 502,166 participants aged 40-69 years. It includes a wide variety of phenotypic data, biological measurements, lifestyle information, detailed genomic data as well as imaging data in a subset of participants.
We obtained data access to all available abdominal DIXON~\cite{Dixon1984-eo} MRI measurements contained in the UKBiobank, totaling to $50,334$ participants.
The only filter criterion applied to the UKBiobank dataset was completion of MRI measurement. 
The DIXON MRI technique used in the UKBiobank scanning protocol provides a full body MRI from neck to knee in six overlapping imaging windows covering a high of $1.10$ meter in total~\cite{slabPaper}. 
Each of these windows consists of four volumetric images covering water, fat, in phase and opposed phase measurements.
After filtering only male subjects, we found $22,911$ participants available for the testis segmentation. 
We convert the provided DICOM files into NIFTI format. 
For analysis, we select the fifth imaging window of dimensions $(224, 162, 72)$ as it contains the testes for most subjects and use the respective water, fat and in phase images for each subject~\cite{slabPaper}. 
Currently, no image alignment for the different imaging windows is applied. 
Details on the UKBiobank image acquisition are outlined in \cref{fig:ukbbmeasurements}.

We found $108$ subjects missing MRI data completely during download from UKBiobank. 
Additionally, in $281$ subjects, at least one file from the fifth imaging window was missing.
Finally, for $23$ subjects the dimension of the image did not adhere to the specification of UKBiobank so that we excluded these patients to provide a consistent dataset.
After this data cleaning procedure we were obtained $22,499$ subjects for subsequent analysis.

\subsection{Annotation}
\begin{figure}
    \centering
    \begin{subfigure}{.3\textwidth}
        \centering
        \includegraphics[width=\textwidth]{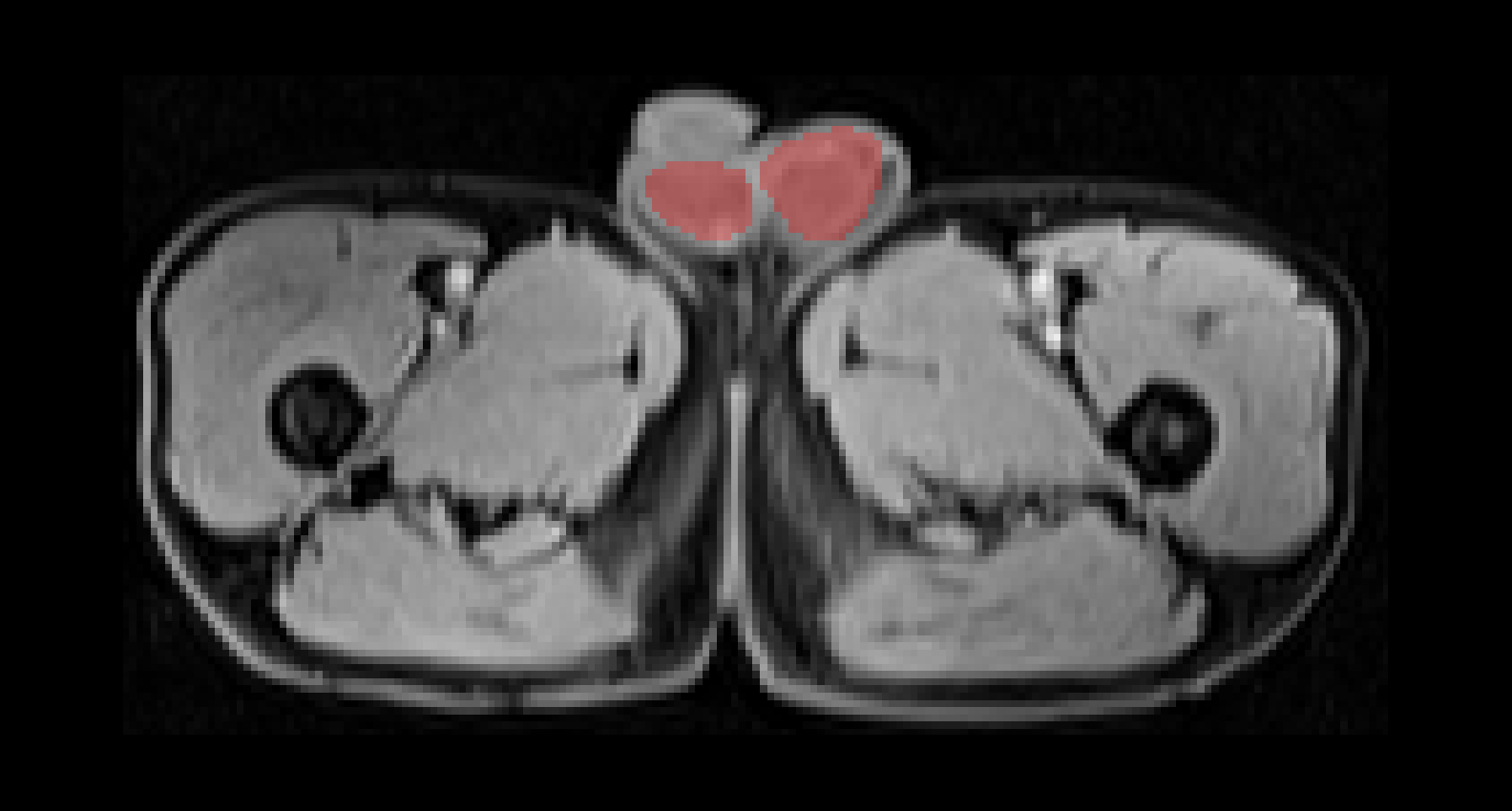}
        \caption{}
        \label{fig:sample_annotations:sfig1}
    \end{subfigure}
    ~
    \begin{subfigure}{.3\textwidth}
        \centering
        \includegraphics[width=\textwidth]{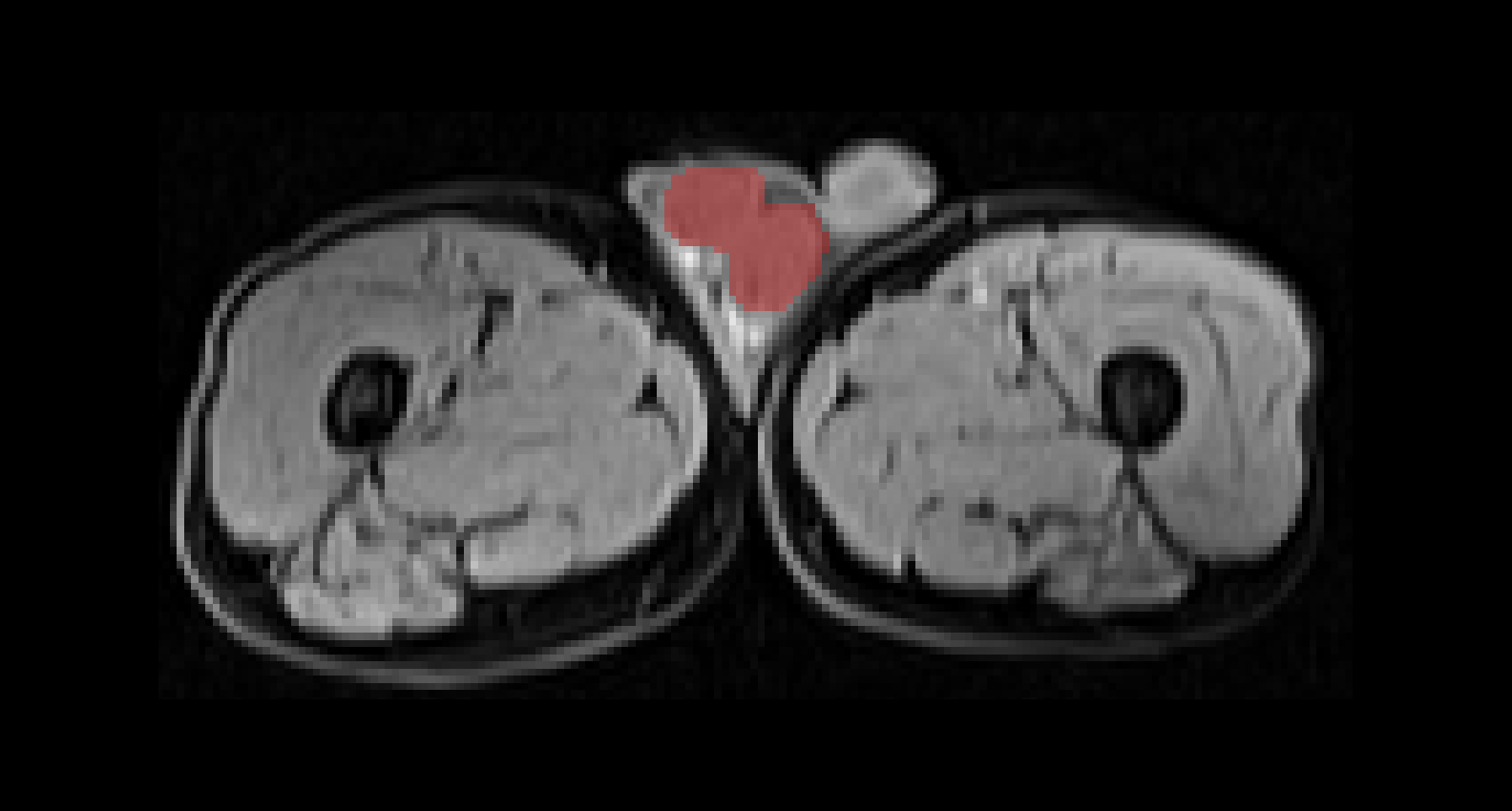}
        \caption{}
        \label{fig:sample_annotations:sfig2}
    \end{subfigure}
    ~
    \begin{subfigure}{.3\textwidth}
        \centering
        \includegraphics[width=\textwidth]{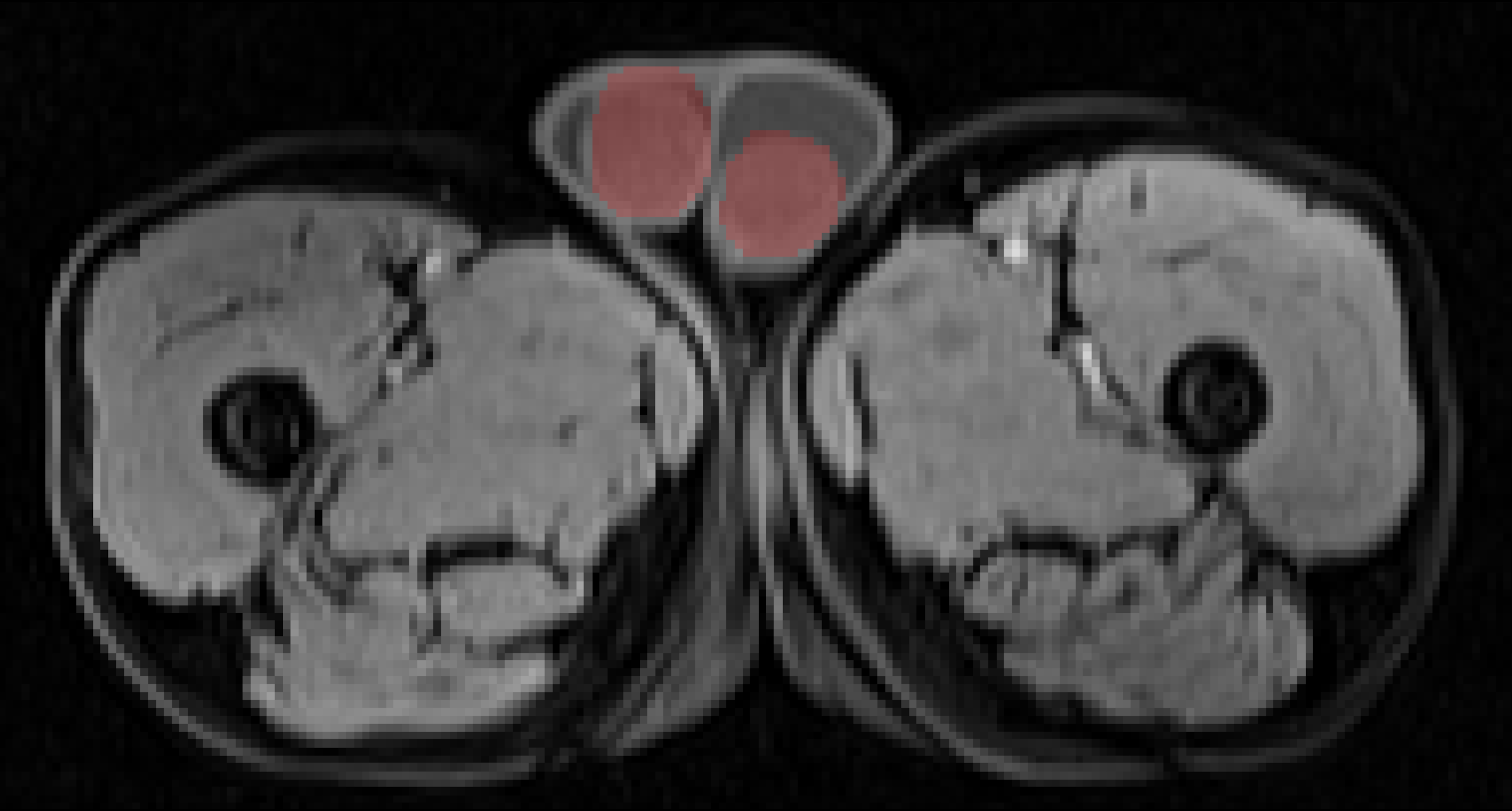}
        \caption{}
        \label{fig:sample_annotations:sfig3}
    \end{subfigure}
    \caption{Sample annotation screenshots from UKBiobank DIXON MRI on the water channel of the fifth imaging window. 
    (a) and (b) are showing MRI scans of physiological testes. 
    (c) shows a hydrocele testis. 
    Reproduced by kind permission of UKBiobank.}
    \label{fig:sample_Annotations}
\end{figure}

\begin{figure*}
    \centering
    \includegraphics[width=.9\linewidth]{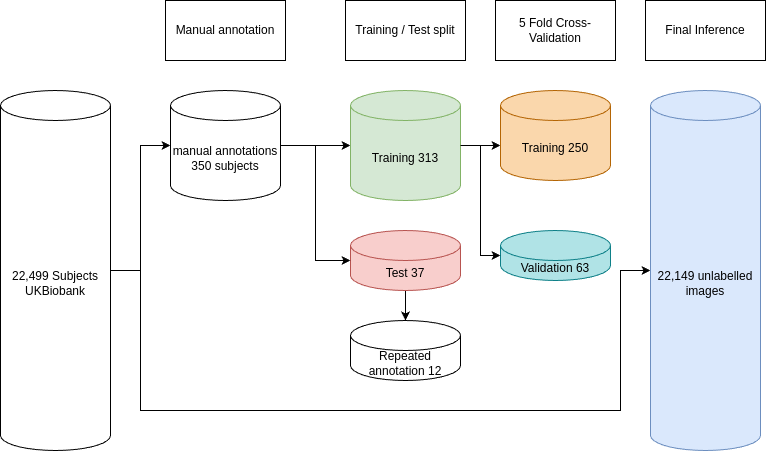}
    \caption{Split of datasets. 
    From the 22,522 initially obtained samples, we annotate 350 randomly selected subjects. 
    Those are split into training and test samples. 
    A subset of 12 images of the testset is then used to calculate interrater agreement. 
    The trainingset is used for 5-Fold Cross-Validation. 
    Final Model is trained on all training subjects (313 subjects) and evaluated on the testset. 
    Final inference is done on 22,172 images.}
    \label{fig:dataset}
\end{figure*}

A randomly sampled subset of the suitable subjects within the UKBiobank was selected as ground truth data. 
These images were manually annotated by a domain expert using a custom annotation software. 
Annotations were obtained for $350$ DIXON MRI images using the water signal of the fifth imaging window. 
Testicles with pathologies, such as hydrocele, were annotated whenever possible, otherwise, they were excluded from the ground truth set and replaced by other scans. 
If testis was not contained or not completely contained in the selected fifth imaging window, the respective subject was not selected for the ground truth set. 
Example annotations are shown in \cref{fig:sample_Annotations}.

We split the resulting dataset in a training set of $313$ images and a testset of $37$ images. 
5-Fold Cross-Validation was performed on the training set only. 
The different sets are visualized in \cref{fig:dataset}.

All annotations were obtained by the main annotator, a trained medical expert. 
For evaluation of annotation consistency, we selected a random subset of $12$ subjects from the test set for one additional annotation per sample by our main annotator. 
This subset is called repeated test (RT) set.
To estimate between-user variances, the RT dataset was also annotated by a trained radiologist.

\subsection{Models}
We evaluated four different 2D architectures and one 3D architecture for testis segmentation on MRI images. 
First, we used DeepLabV3~\cite{deeplabv3} as a baseline. 
Additionally, we trained a DeepLabV3Plus~\cite{deeplabv3plus} and the well-known UNet~\cite{unet} architecture. 
All previously mentioned models were used with a convolutional backbone, namely the ResNet34~\cite{ResNet34} architecture. 
Finally, we evaluated a UNet with a MiT-B0 Segformer~\cite{SegformerMitB1} backbone. 
We used the available implementations of the python-segmentation-models package~\cite{Iakubovskii:2019} with default backbones if not mentioned otherwise and pretrained weights originating from the ImageNet dataset. 

The predictions of the 2D models are finally resampled into 3D volumetric data. 
The model input is consisting of the three selected image types, the water channel, fat and in phase channel for each subject.
Additionally, we included a 3D U-Net model based on the MONAI framework~\cite{monai_consortium_2024_12542217}, which we trained from scratch on the dataset described above.

\subsection{Preprocessing}
After stacking, the respective channels were normalized to match the mean and standard deviation of the ImageNet dataset. 
No further augmentation was employed as first experiments showed no favorable performance after augmentation application. 
We also note that the number of suitable augmentation techniques is limited since commonly applied techniques such as flipping are considered destructive, as left and right testicle are differing in height based on different source of blood supply.  

\subsubsection{2D Models}
In the first step, all models were trained on the annotated ground truth training data using a 5-Fold Cross-Validation. 
We used a batch size of $128$ and predicted two target classes, foreground and background, with Cross Entropy Loss. 
The Adam optimizer~\cite{kingma2017adammethodstochasticoptimization} with default parameters was used. 
The learning rate was set automatically using the pytorch lightning tuner ~\cite{Falcon_PyTorch_Lightning_2019} and early stopping was employed based on the test dice score. 
We calculated dice score for training and test samples. 

We then used the best model of 5-Fold Cross-Validation and fit the respective architecture on all available training data using the test set to apply early stopping. 

\subsubsection{3D Model}
For the 3D U-Net, the training procedure slightly differed from the 2D model setup.
Only the foreground was predicted as a single class and the learning rate was set to $0.001$ and maximum number of epochs was 100.
As a loss function for this model the dice loss~\cite{DBLP:journals/corr/SudreLVOC17} was used and the batch size was adjusted to 12.

\subsection{Postprocessing}
Since not all subjects' testes were entirely contained in the fifth imaging window, we implemented a simple postprocessing to handle the cases of testes at the margin of the imaging window.
We filtered images which contained annotated testis voxels in the volume margins and assigned those images a volume of $0$.
Subjects to which postprocessing was applied were flagged accordingly.

\section{Results}

\subsection{Human level performance}
\begin{figure}
    \centering
        \begin{subfigure}{.3\textwidth}
        \centering
        \includegraphics[width=\textwidth]{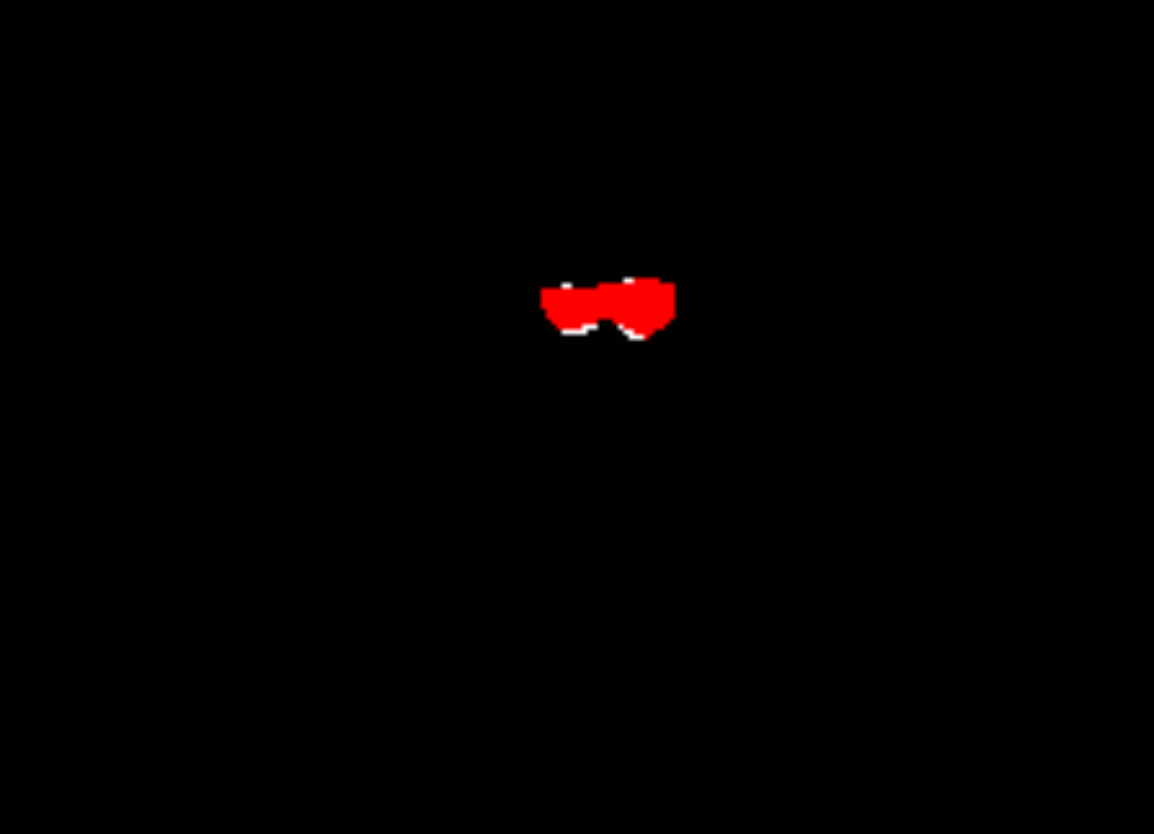}
        \caption{Axial}
        \label{fig:multiple-annotation:sfig1}
    \end{subfigure}
    ~
    \begin{subfigure}{.3\textwidth}
        \centering
        \includegraphics[width=\textwidth]{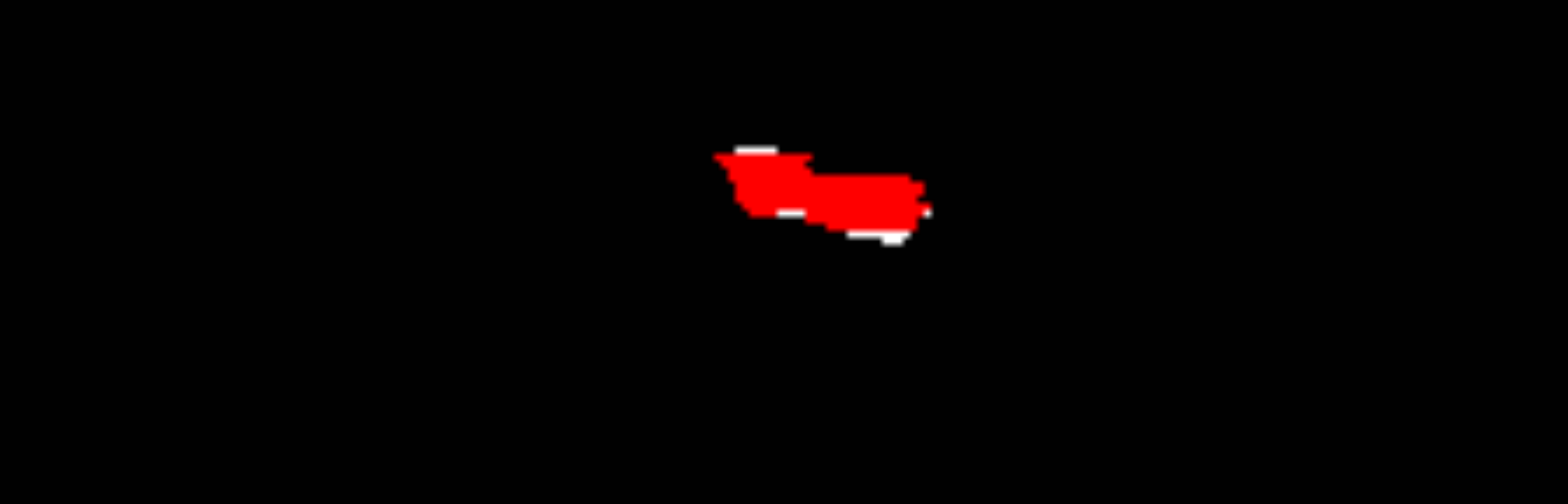}
        \caption{Coronal}
        \label{fig:multiple-annotation:sfig2}
    \end{subfigure}
    ~
    \begin{subfigure}{.3\textwidth}
        \centering
        \includegraphics[width=\textwidth]{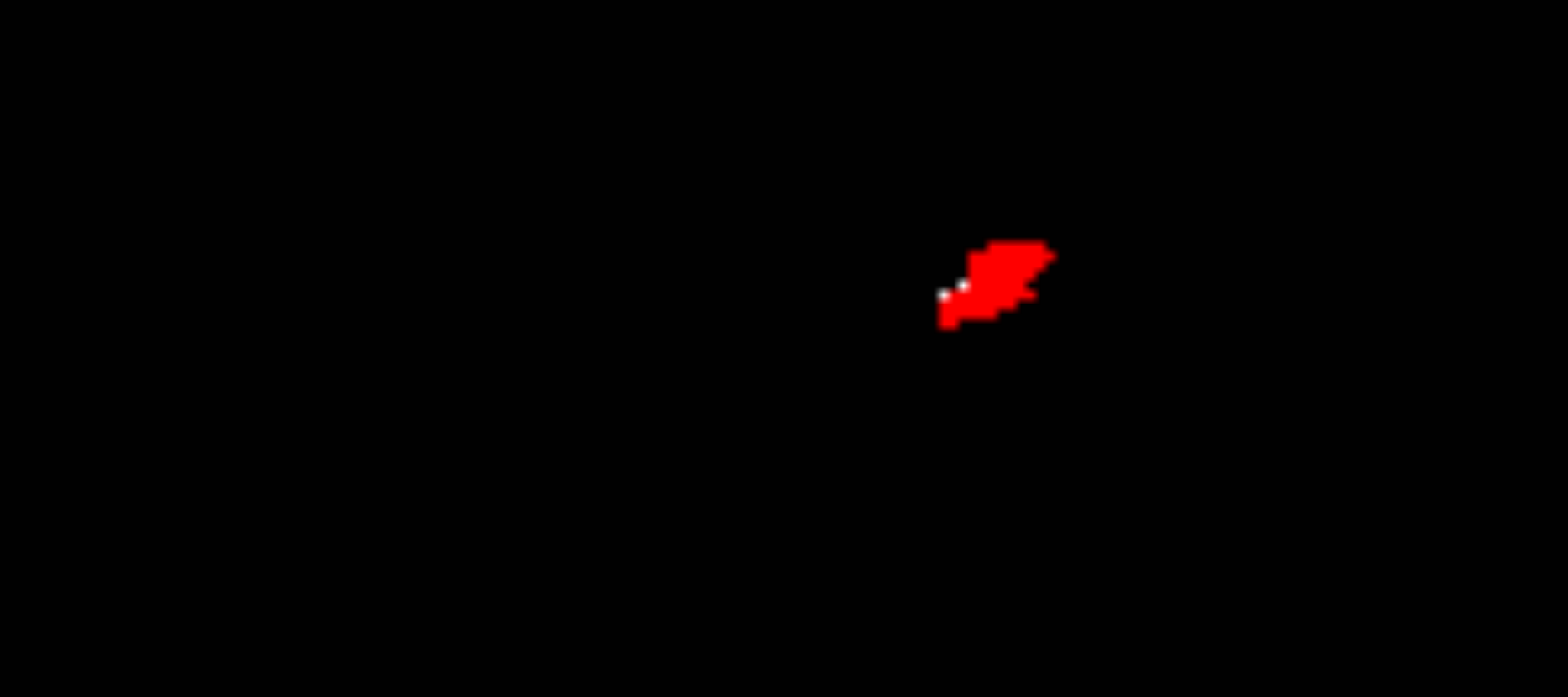}
        \caption{Sagittal}
        \label{fig:multiple-annotation:sfig3}
    \end{subfigure}
    \caption{Multiple annotations for one example image. 
    Annotation from main annotator red, radiologist annotation white. 
    Annotation is viewed as volume in one example view on sagittal, coronal and axial slices.}
    \label{fig:multiple-annotation}
\end{figure}

We compared the dice score between initial annotation and single repeated annotation for the main annotator.
For inter-observer agreement we calculated the dice score between initial annotation of the main annotator and annotation of the trained radiologist.
The median dice score for repeated annotation by the main annotator is $0.90$ (mean $0.90$) and the median dice comparing the annotators is $0.83$ (mean $0.82$). 
One example annotation from both annotators for the same image is shown in \cref{fig:multiple-annotation}.

\subsection{5-Fold Cross-Validation Results}
The results for 5-Fold Cross-Validation are shown in \cref{tab:results}. 
All models reached a dice score around $0.8$. 
The U-Net with ResNet backbone and DeepLabv3+ were the best performing models, with dice $0.85$ and $0.84$ respectively. 
The 3D U-Net yielded a slightly lower performance of $0.83$ compared to the 2D models.
As can be seen the U-Net was considered the best model based on highest mean dice score and lowest standard deviation across 5-Fold Cross-Validation folds. 

\begin{table}[t]
    \centering
    \begin{tabular}{l|c|c}
         Model & Mean & SD \\\hline \hline
         \textbf{U-Net} & \textbf{0.85} & \textbf{0.01}\\\hline
         DeepLabv3 & 0.79 & 0.02\\\hline
         DeepLabv3+ & 0.84 & 0.02 \\\hline
         U-Net transformer & 0.81 & 0.03\\\hline
         3D U-Net & 0.83 & 0.01
    \end{tabular}
    \caption{Dice score mean and standard deviation results for all used models in 5-Fold Cross-Validation. SD: Standard Deviation}
    \label{tab:results}
\end{table}

\subsection{Evaluation of the Final Model}
Due to the superior performance of the U-Net, we retrained this architecture on all available training data by using the (unseen) test set to apply early stopping and report its final dice score. 
After training with the same settings as used in 5-Fold Cross-Validation, we observed a final mean dice score of $0.84$ and median dice of $0.87$ on the test set.

On the RT images, the model obtained a median dice score of $0.87$ (mean $0.84$), showing human level performance in segmenting testis in MRI DIXON data.

\subsection{Quantitative Analysis}
To quantitatively evaluate the results, we calculated the volume of the predicted testis segmentation and compared it to values from literature. 
The volume was determined by calculation of segmentation mask, and accumulation of testis voxel per subject. 
Finally, the number of testis voxel was multiplied with the volume of a single voxel, resulting in cubic millimeter testis volume estimates. 
However, it is important to note that our focus is to minimize technical biases, e.g. inter-observer biases, with regard to volume assessment via orchidometer, as well as selection biases, e.g. inclusion of patients from tertiary referral centres, present in previous studies. 

\subsection{Testis Volume Evaluation}
\begin{figure}
    \centering
    \includegraphics[width=\linewidth]{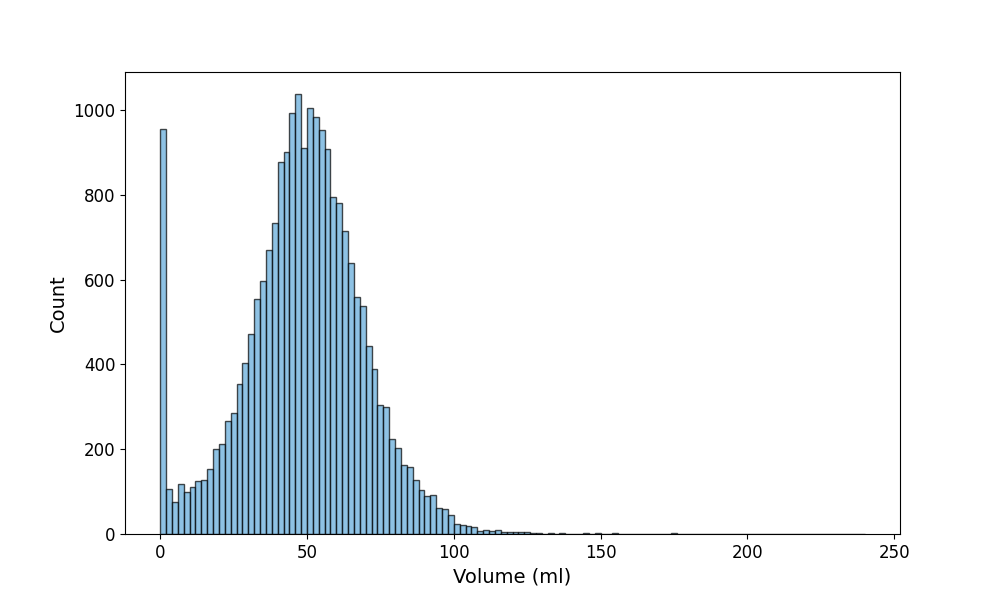}
    \caption{Histogram of calculated volumes for 22,149 unlabeled DIXON MR images from UKBiobank. Count in absolute values and volume in ml.}
    \label{fig:hist}
\end{figure}
The results of the volume calculation for the previously unseen images is shown in \cref{fig:hist}.
The testis volumes of subjects within the normal distribution range appeared reasonable and fell within expected ranges based on existing literature based on the current gold standard, i.e. ultrasonography~\cite{Tuttelmann2012-xe,Hansen2023-js}.
This indicates that the model performs well for the majority of subjects, providing reliable segmentation results for typical testis sizes. 

A fraction of $7.4\%$ ($>+2$SD: $5.2\%$, $<-2$SD: $2.2\%$) of the dataset shows testis volumes outside 2 standard deviations ($2$SD) of the mean. The mean bi-testicular volume is $48.5$ mL (SD: $21.3$ mL, $-2$SD: $6.0$ mL, $+2$SD: $91.1$ mL).

$595$ subjects were assigned a $0$ volume annotation during postprocessing.

\section{Conclusion}
For the first time, we provide a model capable of testis segmentation using MRI data based on a population based dataset. 

Overcoming limitations of conventional methods as well as avoiding bias of MRI segmentation approaches based on patient-based data for the assessment of testicular volume, our approach allows novel study approaches based on the comprehensive data of UKBiobank participants. 
This includes population-based genetic approaches on male fertility that have previously been hampered by under reported infertility diagnoses~\cite{Venkatesh2024.03.19.24304530}.

The task of testis segmentation is inherently difficult on the DIXON MRI data contained in the UKBiobank, due to a low resolution of the MRIs, making it difficult to differentiate between testicular and surrounding tissue.
Additionally, the mobility of the testes further complicates the segmentation process, as certain positions make it more difficult to differentiate testicular tissue from other tissues.
This is also reflected in the low Dice scores obtained by the human experts. 
Using the UKBiobank as data source, we are able to estimate testis volume at population scale for the first time.

While the model exhibits reasonable performance for the vast majority of subjects in UKBiobank, some predictions are by far too small or large. 
For subjects with significantly smaller predicted bi-testicular volumes (below $2$ ml), the scans often show technical abnormalities, making accurate segmentation more challenging. 
This included instances of testes positioned at the margins of the MRI imaging field which occurred frequent as the MRI imaging field of the UKBiobanks imaging protocol had not been optimized for testicular imaging. 
Alignment of the overlapping imaging windows might mitigate these effects, however potentially introducing novel sources of error.
As alignments can be error-prone and this can lead to additional errors and exclusion criteria.
We therefore decided to not use alignments in our study. 

Subjects with significantly higher predicted bi-testicular volume often exhibit distinct testicular pathologies, e.g. hydrocele. 
The model has not specifically trained to identify these rare pathologies.
This suggests that a specialized model for detecting these pathologies, or a general model incorporating a broader range of pathological data, could be developed to improve accuracy. 
It is important to note that in other cases the segmentation was correct despite testicular pathologies, which supports the technical feasibility of implementing a more specialized model.
However, developing such models is beyond the scope of the current analysis.
Specialized models for identification of Pathology would help to identify those subjects and will be investigated in future work based on our established pipeline.

Evaluation of more advanced recent transformer models and additional 3D architectures will be one in future work.
Especially interesting would be integration of currently identified special cases, like hydrocele mentioned above. 
Given the reasonable predictions from this initial model and manual refined labels for edge cases will provide a novel dataset ready for training of more advanced architectures.
Notably, our approach also enables a human in the loop approach by iteratively generating segmentations for unseen subjects, manually refinement and novel training.
This approach is also to be evaluated in future work.

Our pipeline introduces means to conduct population scale analysis of testis volume effects for the first time, enabling population-based analysis on male fertility.
\section{Acknowledgments}
A.S.B. is funded by the Deutsche Forschungsgemeinschaft (German Research Foundation), no. 64240267.

{\small
\bibliographystyle{ieee_fullname}
\bibliography{main}
}

\end{document}